\documentclass[runningheads,fleqn]{svmult}

\usepackage{makeidx}   % allows index generation
\usepackage{graphicx}  % standard LaTeX graphics tool
\usepackage{subeqnar}  % subnumbers individual equations
\usepackage{multicol}  % used for the two-column index
\usepackage{physproc}  % flushleft layout of diverse elements,etc.
\makeindex             % used for the subject index
\begin{document}
\title*{Monte Carlo Simulations of Quantum Spin Systems in the Valence Bond Basis}
\toctitle{Monte Carlo Simulations of Quantum Spin Systems in the Valence Bond Basis}
% allows explicit linebreak for the table of content
%
%
\titlerunning{Monte Carlo Simulations of Quantum Spin Systems}
% allows abbreviation of title, if the full title is too long
% to fit in the running head
%
\author{Anders W. Sandvik\inst{1}
\and K. S. D. Beach\inst{1,2}}
\authorrunning{A. W. Sandvik and K. S. D. Beach}
% if there are more than two authors,
% please abbreviate author list for running head
%
%
\institute{Department of Physics, Boston University,
Boston, Massachusetts, USA
\and Theoretische Physik I, Universit\"at W\"urzburg,  W\"urzburg, Germany}

\maketitle              % typesets the title of the contribution

\begin{abstract}
\index{abstract}
We discuss a projector Monte Carlo method for quantum spin models formulated in 
the valence bond basis, using the $S=1/2$ Heisenberg antiferromagnet as an example. 
Its singlet ground state can be projected out of an arbitrary basis state as the trial 
state, but a more rapid convergence can be obtained using a good variational state. As an 
alternative to first carrying out a time consuming variational Monte Carlo calculation, 
we show that a very good trial state can be generated in an iterative fashion in the course 
of the simulation itself. We also show how the properties of the valence bond basis enable 
calculations of quantities that are difficult to obtain with the standard basis of $S^z$ 
eigenstates. In particular, we discuss quantities involving finite-momentum states in the 
triplet sector, such as the dispersion relation and the spectral weight of the lowest triplet.

\end{abstract}

\section{Introduction}

Quantum Monte Carlo (QMC) simulations of spin systems have traditionally been carried out
in the basis of eigenstates of the spin-$z$ operators $S^z_i$, $i=1,\ldots, N$, i.e., the basis 
of ``up" and ``down" spins in the case of $S=1/2$ (which is the case we consider here). 
For the prototypical model of interacting quantum spins, the antiferromagnetic ($J>0$) Heisenberg 
hamiltonian,
\begin{equation}
H = J\sum_{\langle i,j\rangle} {\bf S}_i \cdot {\bf S}_j =
 J\sum_{\langle i,j\rangle} [S^z_iS^z_j + \hbox{$\frac{1}{2}$}(S^+_iS^-_j+S^-_iS^+_j)],
\label{heisenberg}
\end{equation}
this basis is clearly natural and convenient, as an off-diagonal operator 
acting on a basis state just flips two spins or destroys the state. 
Starting with the work of Suzuki \cite{suzuki}, finite-temperature simulation methods 
employing the spin-$z$ basis were developed in which a quantum mechanical expectation value for a 
system in $D$ dimensions is mapped onto an anisotropic classical statistical-mechanics problem 
in $D+1$ dimensions---the discretized \cite{barma,hirsch,cullen} or continuous \cite{beard,prokofev} 
imaginary-time path integral. There are now very efficient methods utilizing loop-cluster 
\cite{evertz1,evertz2,beard} or ``worm" \cite{prokofev} updates of the world-line 
spin configurations. These methods have enabled studies of systems with $\approx 10^4-10^5$ spins in 
the low-temperature (ground-state) limit and much more at elevated temperatures. Loop updates have 
been developed \cite{sandvik1,syljuasen} also for the alternative and now frequently 
used power-series expansion representation \cite{handscomb,lee84,sandvik2,sandvik3} of the partition 
function (stochastic series expansion; SSE), where the spin-$z$ basis is also normally used. 
It is in principle possible to adapt these approaches to other local bases, e.g., that of 
singlet and triplet states of spin pairs on a dimerized lattice. This basis is often used
in diagrammatic and series-expansion calculations \cite{valeri}, but its implementation in 
QMC simulations is typically rather cumbersome.

\begin{figure}[t]
\centerline{\includegraphics[width=6.75cm]{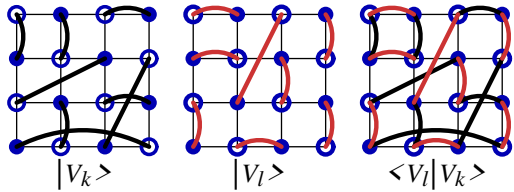}}
\caption{Two valence bond states $|V_k\rangle$, $|V_l\rangle$ in two dimensions
and their overlap graph corresponding to $\langle V_l|V_k\rangle=2^{N_\circ - N/2}$, 
where $N_\circ$ is the number of loops formed (in this case $N_\circ=3$ and the number of 
sites $N=16$). Filled and open circles correspond to sublattices $A$ and $B$. The sign convention 
in Eq~(\ref{singletbond}) for a singlet valence bond $(i,j)$ dictates that spins $i$ and $j$ 
belong to $A$ and $B$, respectively} 
\label{vbloops}
\end{figure}

Zero-temperature ($T=0$) simulations, in which the ground state is projected out of a trial wave
function, are also normally carried out in the spin-$z$ basis \cite{trivedi,sorella}. Here we 
will discuss an alternative approach to ground-state calculations which turns out to have some unique 
features enabling access to quantities that are normally difficult to obtain with standard 
finite-temperature or projector methods. We make use of the valence bond basis, i.e., states 
in which the spins are paired up into singlets;
\begin{equation}
|V \rangle = |(i_1,j_1)(i_2,j_2)\cdots (i_{N/2},j_{N/2})\rangle .
\end{equation}
Here $(i,j)$ denotes a singlet formed by the spins at sites $i$ and $j$;
\begin{equation}
(i,j)=(|\uparrow_i\downarrow_j\rangle-|\downarrow_i\uparrow_j\rangle)/\sqrt{2},
\label{singletbond}
\end{equation}
and the total number of sites $N$ is assumed to be even. While in principle
one can include all possible pairings of the spins, it is in most cases better to 
consider a smaller basis in which the sites are first divided into two groups, A and B, 
of $N/2$ spins each, and to only consider singlets $(i,j)$ in which the first index $i \in A$ and 
the second $j \in B$ \cite{hul38,sut88,lia88}. In the case of a bipartite lattice, these groups are 
naturally the two sublattices, as shown in Fig.~\ref{vbloops}. This restricted VB basis has $(N/2)!$ 
states and is still massively overcomplete---the singlet space has $N!/[(N/2)!]^2(N/2+1)$ dimensions 
\cite{hul38}. The VB basis states are all non-orthogonal, overlapping with each other according to 
the simple loop rule illustrated in Fig.~\ref{vbloops}.

The VB basis was introduced already in the early 1930s \cite{hul38,rum32,pau33} and has 
played an important role in exactly solvable models \cite{hul38,maj69,ss81,aklt}. Later, it 
became a tool for describing spin liquids---the resonating valence bond (RVB) 
mechanism introduced by Fezekas and Anderson \cite{fez74,and87,fradkinbook}, in which the ground 
state is dominated by short valence bonds. In exact diagonalization studies, the VB basis is useful 
in cases where it is a good approximation to only consider a restricted (and incomplete) space of 
short bonds (spin liquids and other states with no magnetic long-range order)
\cite{iske,kohmoto,tang,mambrini}. Variational calculations in the VB basis have been carried 
out for the 2D Heisenberg model \cite{lia88,jielou}. Furthermore, Liang realized that a variational 
VB state could be considerably improved by stochastically projecting it with an operator $(-H)^m$ 
for large $m$ \cite{lia90}. Later, Santoro et al.~devised a Green's function method for calculating 
energies in the VB basis \cite{san99}. Despite the promising results obtained in these studies, 
there was, to our knowledge, no further developments of QMC methods in the VB 
basis until one of us recently introduced two related projector algorithms \cite{sandvikvb}, improving on 
the schemes of Liang \cite{lia90} and Santoro et al.~\cite{san99}. These algorithms have already 
been applied in studies of quantum phase transitions \cite{jq2d,jq3d} and entanglement 
entropy \cite{alet}. 

Some previously unnoticed advantages of the VB basis in QMC algorithms were 
pointed out in Ref.~\cite{sandvikvb}. Here we summarize our recent work on VB projector methods 
and highlight some of their unique features. We discuss in particular a scheme for 
``self-optimizing" the trial state out of which the ground state is projected, and also show
how to study properties of triplet excitations at finite momentum. 

\section{Ground state projection}

Consider a state $|\Psi\rangle$ and its expansion in terms of eigenstates
$|n\rangle$, $n=0,1,\ldots$, of some hamiltonian $H$;
\begin{equation}
|\Psi\rangle = \sum_n c_n |n\rangle.
\end{equation}
With $C$ a constant chosen such that the lowest eigenvalue $E_0-C$ is the largest in magnitude, 
a large number $m$ of repeated operations with $C-H$ projects out the ground state,
\begin{equation}
(C-H)^m |\Psi\rangle \to c_0(C-E_0)^m \left [ |0\rangle + 
\frac{c_1}{c_0}\left (\frac{C-E_1}{C-E_0} \right )^m |1\rangle +\ldots \right ],
\label{projection}
\end{equation}
provided that the overlap $c_0 \not=0$. Here we will first consider singlet eigenstates 
of the Heisenberg model (\ref{heisenberg}), which can be expanded in VB states;
\begin{equation}
|\Psi \rangle = \sum_i f_i |V_i\rangle .
\end{equation}
Because of the overcompleteness of the VB basis, this expansion is not unique. That, however, 
does not prohibit the ground state $|0\rangle$ to be projected out according to Eq.~(\ref{projection}). 
The Heisenberg hamiltonian can be written in terms of singlet projection operators 
$H_b \equiv H_{i(b),j(b)}$ on the 
interacting spin pairs, $\{i(b),j(b)\}$, $b=1,\ldots ,N_b=DN$ (for a periodic
cubic $D$-dimensional lattice);
\begin{equation}
H= -J\sum_{b=1}^{N_b} H_{b} = 
-J\sum_{b=1}^{N_b} H_{i(b),j(b)},~~~~ H_{ij}=-(\hbox{$\frac{1}{4}$}-{\bf S}_i \cdot {\bf S}_j).
\end{equation}
When a singlet projector $H_{ij}$ acts on a VB basis state, one of two things
can happen; 1) if $i,j$ belong to the same bond the state is unchanged with a matrix element $1$, 
or 2) if they belong to different bonds these two bonds are reconfigured (``flipped") with matrix 
element $1/2$;
\begin{eqnarray}
& & H_{ij}|\cdots (i,j)\cdots \rangle =|\cdots (i,j)\cdots \rangle, \label{hijdia} \\
& & H_{ij}|\cdots (i,k)\cdots (l,j)\cdots \rangle =
\hbox{$\frac{1}{2}$}|\cdots (i,j)\cdots (l,k)\cdots \rangle. \label{hijoff}
\end{eqnarray}
Here positive-definitness of (\ref{hijoff}) is directly related to the two sites $i$ and $j$ 
being in different sublattices. For a frustrated system, where there are operators with both sites $i,j$ 
in the same sublattice, positive-definitness does not hold \cite{sut88,lia88}. For a non-frustrated system 
the simple bond flip (\ref{hijoff}) makes for a convenient stochastic 
implementation of the ground state projection (\ref{projection}). We write the projection operator as
(with $J=1$ henceforth)
\begin{equation}
(C-H)^m = \left (\sum_{b=1}^{N_b} H_b \right)^m 
= ~\sum_r P_r,~~~~~ P_r =  H_{b^r_{N/2}}\cdots H_{b^r_{2}}H_{b^r_{1}},
\end{equation}
where we have introduced a compact notation $P_r$, $r=1,\ldots,N_b^m$, for the different 
strings of singlet projectors. When a string $P_r$ acts on a VB basis state 
$|V\rangle$ the result is another basis state, which we denote $|V(r)\rangle$, with 
a prefactor (weight) $W_{r}$ which is simply given by the number $m_{\rm off}$ of 
off-diagonal operations in the course of evolving $|V\rangle$ to $|V(r)\rangle$;
\begin{equation}
P_r |V\rangle = W_{r}|V(r)\rangle,~~~~ W_{r}=2^{-m_{\rm off}}.
\label{wrk}
\end{equation}

\begin{figure}[t]
\centerline{\includegraphics[width=10.5cm]{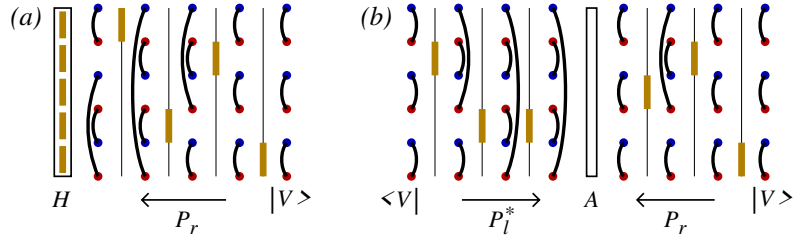}}
\caption{Propagation of a VB state on a 6-site chain. The horizontal bars represent
nearest-neighbor Heisenberg interactions (singlet projectors). In the single projection (a) 
the state is propagated from right to left, and an estimator for the ground state energy is obtained 
by acting once more with all terms of the hamiltonian. In the double projection (b) the state is 
projected from the right and the left, and any operator expectation value can be estimated by 
calculating the corresponding matrix elements between the propagated states}
\label{propagation}
\end{figure}

We here first consider projecting the ground state out of a single VB basis state; later
we will consider the use of a more complicated trial state. We consider two ways 
to calculate expectation values:
\begin{eqnarray}
\label{hexp}
\langle H\rangle & = & 
\frac{\sum_r \langle \Psi | H P_r |V\rangle}{\sum_r \langle \Psi | P_r |V\rangle} =
\frac{\sum_r W_{r} \langle \Psi | H |V(r)\rangle}{\sum_r W_{r} \langle \Psi |V(r)\rangle}, \\
\label{aexp}
\langle A\rangle & = & \frac
{\sum_{rl} \langle V | P_l^* A P_r |V\rangle}
{\sum_{rl} \langle V | P^*_l P_r |V\rangle}=
\frac{\sum_{rl} W_{r}W_{l}\langle V(l) | A |V(r)\rangle}
{\sum_{rl} W_{r}W_{l} \langle V(l)  |V(r)\rangle}.
\end{eqnarray}
We will discuss how to estimate these using importance sampling; terms 
(configurations) of the numerators are illustrated in Fig.~\ref{propagation}. We will refer to 
(\ref{hexp}) and (\ref{aexp}) as the single and double projection, respectively.

In (\ref{hexp}), which is an exact (when $m\to \infty$) expectation value only of the 
hamiltonian  (or other operators for which the ground state is an eigenstate) 
the state $|\Psi\rangle$ is in principle arbitrary. It is very convenient to use a 
state which has equal overlaps with all the VB basis states, e.g., the N\'eel state $|\Psi_N\rangle$
(all spins up on sublattice A and down on B). It is easy to see that 
$\langle \Psi_N |V\rangle = 2^{-N/2}$ for any basis state $|V\rangle$. Since $H |V(r)\rangle = 
-\sum_b H_b |V(r)\rangle$ is a sum of basis states multiplied by factors $-1$ or $-1/2$,
the overlaps with $\langle \Psi_N|$  drop out altogether and do not have to 
be considered further. If the projector 
strings $P_r$ in (\ref{hexp}) are importance-sampled according to their weights $W_{r}$, 
the estimator for the ground state energy is thus
\begin{equation}
E_0 = \langle H\rangle = -\langle m_{\rm d} + \hbox{$\frac{1}{2}$} m_{\rm o}\rangle,
\end{equation}
where $m_{\rm d}$ and $m_{\rm o}$ are, respectively, the number of diagonal and off-diagonal
operations $H_b |V(r)\rangle$ (and $m_{\rm d}+m_{\rm o}=N_b$). It should be noted that
although this estimator is exact in the limit $m \to \infty$, it is not variational. The correct 
energy may thus be approached with increasing $m$ from above or below.

Eq.~(\ref{aexp}) is valid for any expectation value and in the case of $A=H$ gives a variational 
estimate of the energy. Using $W_{r}W_{l}\langle V(l)|V(r)\rangle$ as the sampling weight, 
the estimator for any $\langle A\rangle$ is of the form
\begin{equation}
\langle A\rangle =
\left \langle \frac{\langle V(l) | A |V(r)\rangle}{\langle V(l) |V(r)\rangle}\right\rangle.
\label{esingle}
\end{equation}
In the case of a spin correlation function $\langle {\bf S}_i \cdot {\bf S}_j\rangle$, the
matrix element is related to the loop structure of the overlap graph \cite{sut88,lia88}
(illustrated in Fig.~\ref{vbloops}):
\begin{equation}
\frac{\langle V_l | {\bf S}_i \cdot {\bf S}_j |V_r\rangle}{\langle V_l |V_r\rangle} =
\left\lbrace
\begin{array}{ll}
+3/4, &~\hbox{\rm if $i,j \in$ same loop, same sublattice,} \cr
-3/4, &~\hbox{\rm if $i,j \in$ same loop, different sublattices,} \cr
~~~0, &~\hbox{\rm if $i,j \in$ different loops.}
\end{array}\right.
\end{equation}
Measuring the spin correlations is hence straight-forward once the overlap-loops have 
been constructed. Higher-order functions, e.g., dimer-dimer correlations 
$\langle ({\bf S}_i \cdot {\bf S}_j)({\bf S}_k \cdot {\bf S}_l)\rangle$, are also
related to the loop structure \cite{kevinvb}.

Note again that no bond operator
$H_b$ can destroy a VB state and that all the states have non-zero overlap with each other. Thus all 
terms in (\ref{aexp}) contribute to the expectation value. This turns out to be an advantage in 
constructing a Monte Carlo algorithm, as any change made in the operator strings can
be accepted with some probability. With an orthogonal basis, such as the $S^z_i$ eigenstates, there
would be considerable constraints, both in terms of individual operators in the projection 
[the spin flip operator in (\ref{heisenberg}) can act, without destroying the state, 
only on anti-parallel spins], and in ensuring a non-zero overlap between the propagated states 
(the two propagated states have to be identical). Note also that the singlet projectors
are non-hermitian in the VB basis. As indicated in (\ref{aexp}), and illustrated
in Fig.~\ref{propagation}, we here propagate two states, $|V(l)\rangle \propto P_l|V\rangle$ 
and $|V(r)\rangle \propto P_r|V\rangle$, and subsequently compute their overlap and various 
matrix elements. Propagating $|V\rangle$ with $P^*_lP_r$ and then taking the overlap with 
$|V\rangle$ is not equivalent term-by-term.

To carry out the projection stochastically, the operator strings are stored in arrays 
$[P_\alpha ] =  [b^\alpha_1][b^\alpha_2]\cdots [b^\alpha_m]$, where $b^\alpha_i \in \{ 1,\ldots,N_b \}$ 
with $\alpha=1,2$ corresponding to $P_r$ and $P_l$, respectively, in the double projection; 
in the single projection $\alpha$ is redundant. A table holds the site pairs $i(b),j(b)$. 
The state $|V\rangle$ is stored in a list $[V] = [v_1][v_2]\cdots [v_N]$ where $v_i=j$ and 
$v_j=i$ if there is a valence bond at $(i,j)$. Propagation with the bond flips (\ref{hijoff}) is easily carried 
out in this representation. The state list $[V]$ is then first copied into two lists, $[V_1]$ and $[V_2]$, 
in which $|V(r)\rangle$ and $|V(l)\rangle$ are constructed.

The simulation can be started with a randomly generated operator string. The strings 
can can be updated in an trivial way, by changing a number $R$ of operators at random. 
In either $[P_1]$ or $[P_2]$, $R$ positions $p_j$, $j=1,\ldots,R$, $p_j \in \{1,\ldots,m \}$ 
(all different) are generated. Their contents $b^\alpha_{p_j}$ are picked randomly from the set 
$\{1,\ldots ,N_b\}$ (excluding the old value for each $p_j$). To calculate the 
Metropolis acceptance probability, the state is propagated with the updated operator
string and the number of off-diagonal operations $m_{\rm off}$ in (\ref{wrk}) is counted.
In the single expansion, the acceptance probability is simply
\begin{equation}
P_{\rm accept} = {\rm min}\left [ \frac{W_{r}^{\rm new}}{W_{r}^{\rm old}},1\right ]
={\rm min}[2^{m_{\rm off}^{\rm old}-m_{\rm off}^{\rm new}},1],
\label{pacc}
\end{equation}
whereas in the double projection an overlap ratio appears as well. 
In the double projection, we change operators only in one of the 
operator strings at a time, so that only one state has to be propagated. It is customary 
to define a size-normalized Monte Carlo step (or ``sweep"). For projector length $m$ 
we do $m$ replacement attempts, and so our sweep is independent of the number of operators $R$
replaced in each update.

Normally, in Monte Carlo simulations one does not compute the full weight, because it
is possible to more speedily calculate just the change in the weight [the weight ratio in (\ref{pacc})]. 
In the present formulation of the VB projector algorithm the weight is, however, recomputed from 
scratch each time, because a better scheme is not yet known. Each update hence requires on the order 
of $m$ operations. In the double expansion, construction of the loops needed to compute the overlap 
scales as $N$, but typically $m > N$ and the propagation of the state dominates the simulation. 
In spite of the need to recalculate the weight, the scheme is sufficiently efficient to compete with 
other ground state QMC methods. More importantly, as we will discuss in Sec.~4, the VB basis 
offers access to quantities out of reach for other methods.

\begin{figure}[t]
\centerline{\includegraphics[width=9.75cm, clip]{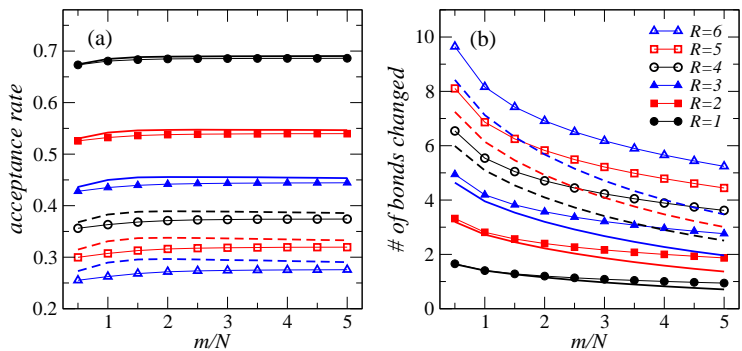}}
\caption{(a) Acceptance rate in a double projection and (b) the number of bonds changed in the 
propagated state in accepted updates for a $16\times 16$ lattice (symbols with lines; the bare lines
are for $L=8$) versus the projection length (normalized by the number of sites $N$). 
The trial state $|V\rangle$ was a columnar dimer state}
\label{arate}
\end{figure}

The optimum number $R$ of operators to replace depends on the acceptance rate.
In Fig.~\ref{arate}(a) we show the acceptance rate for a double-projected 2D system
versus the length $m$ of the projection for $R=1-6$. As expected, the acceptance rate decreases with 
increasing $R$, but it does not change appreciably with $m$. It also depends only 
weakly on the lattice size. Multiplying the acceptance rate by $R$ gives the average number of operators 
changed; it initially increases with $R$ but has a maximum for $R=7$ (for $L=16$). 
The optimum $R$ is clearly model/lattice dependent. Another characteristic of the update is 
the number of bonds changed in the projected state $|V(r)\rangle$ as a consequence of the 
modifications of $P_r$. This number is shown in Fig.~\ref{arate}(b). It is 
seen to increase with $R$, as expected. As a function of $m$ the number of changed bonds 
decreases. This behavior reflects a loop structure of the singlet-projection 
operators \cite{aiz94,evertz2}, which implies that some changes ``upstream" in the operator string 
may be healed  further downstream in the propagation. For a finite lattice in the limit $m \to \infty$ 
one would expect a substitution of an operator far upstream in $P_r$ to have no effect on the final 
propagated state $|V(r)\rangle$. This does not imply that this update is inconsequential, as the 
sampling is over paths, not just the final states in the propagations.
In principle, it should be possible to take advantage of the underlying loop structure of
the singlet projectors \cite{aiz94,evertz2} to devise a loop update in the VB basis, analogous 
to such updates in world-line \cite{evertz1,evertz2} and SSE \cite{sandvik1} simulations. However, 
we have not yet been able to construct a scheme that is in practice faster than the trivial random 
substitution with full state propagation.

\section{Self-optimized trial state}

So far, we have projected the ground state out of a single VB basis state $|V\rangle$. 
This works well \cite{sandvikvb}, but the rate of convergence with $m$ of course depends on 
the state chosen; ideally one would like to maximize the overlap $\langle V|0\rangle$. 
One way to obtain a typically good single-configuration trial state is to first start with an 
arbitrary one; a regular bond pattern or a randomly generated configuration. After carrying 
out some projection steps with this state, the current propagated state $|V(r)\rangle$ 
is chosen as a new trial state. Since this state has been generated in the projection it should
contribute substantially to the ground state and hence typically will be better than a 
completely arbitrary one. However, as we will discuss next, we can do much better than this.

\begin{figure}[t]
\centerline{\includegraphics[width=9cm]{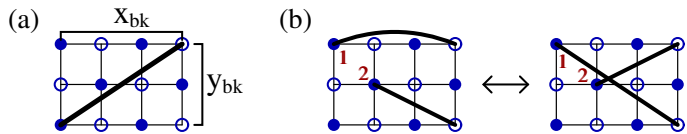}}
\caption{(a) Definition of the size of a bond $b$. (b) Reconfiguration of two bonds in an update 
of the trial state. The two sites marked $1,2$ are chosen at random among all pairs of next-nearest 
neighbor sites.}
\label{xybond}
\end{figure}

Liang's original motivation for introducing a projector technique in the VB basis was to
improve on a variational calculation \cite{lia90}. Previously, Liang, Doucot, and Anderson 
had studied a variational amplitude-product state for the 2D Heisenberg model 
of the form \cite{lia88}
\begin{equation}
|\Psi\rangle = \sum_k f_k |V_k\rangle,~~~f_k = \prod_{b=1}^{N/2} h(x_{bk},y_{bk}),
\label{psihprod}
\end{equation}
where $x_{bk}$ and $y_{bk}$ are the $x$- and $y$-lengths of bond $b$ in VB state $k$, as
illustrated in Fig.~\ref{xybond}(a). Liang et al.~tried power-law and exponential forms depending 
only on the total length $l$ of the bonds [the "Manhattan" length $l=x+y$ was used, but defining 
$l=(x^2+y^2)^{1/2}$ should not change things qualitatively], in addition to keeping several short-bond 
amplitudes as parameters to optimize \cite{lia88}. More recently, all the amplitudes were optimized 
without any assumed form on lattices with up to $32\times 32$ sites, with the result that 
$h(l) \sim l^{-3}$ for long bonds \cite{jielou}. 
One of us has also recently showed the more general result 
$h(l) \sim l^{-(D+1)}$ within a mean-field approach for a $D$-dimensional 
cubic lattice \cite{kevinmf}. In 2D the fully-optimized amplitude-product state turns out to be 
extremely good, with an energy deviating by only $\approx 0.06\%$ from the exact ground state energy 
and with the long-distance spin-spin correlations reproduced to within $2\%$ \cite{jielou}.

With the state (\ref{psihprod}) an expectation value is given by
\begin{equation}
\langle A\rangle = \frac{\sum_{kp} f_kf_p \langle V_p|V_k\rangle
\frac{\langle V_p|A|V_k\rangle}{\langle V_p|V_k\rangle}}
{\sum_{rl} f_kf_p \langle V_p|V_k\rangle},
\end{equation}
which can be evaluated using importance sampling of the VB configurations with weight 
$f_kf_p \langle V_p|V_k\rangle$. Liang et al. introduced a very simple scheme for updating the dimer 
configurations \cite{lia88}, which we here illustrate in Fig.~\ref{xybond}(b). 
Choosing two next-nearest-neighbor sites, i.e.,
ones on a diagonal of a 4-site plaquette (or, in principle, any two sites on the same sublattice),
the two bonds connected to them are reconfigured in the only possible way which maintains only
bonds between the A and B sublattices, as shown in the figure. Labeling the two initially
chosen sites $1$ and $2$, and the bonds connected to them $b=1,2$, the  Metropolis acceptance
probability is, assuming that the bond update was made in $|V_k\rangle$, resulting in $|V_{k'}\rangle$,
\begin{equation}
P_{\rm accept} = {\rm min}\left [
\frac{h(x_{1k'},y_{1k'})h(x_{2k'},y_{2k'})}
{h(x_{1k},y_{1k})h(x_{2k},y_{2k})}
\frac{\langle V_p| V_{k'}\rangle}{\langle V_p| V_k\rangle},1\right ].
\label{hratio}
\end{equation}

We can use an amplitude-product state as the trial state in the projector QMC method, using 
some set of amplitudes not necessarily originating from a variational calculation. In updating 
the bond configurations, we then also must compute the new weight of the propagation; 
$P_r|V_{k'}\rangle = W_{k'r}|V_{k'}(r)\rangle$, and of course $\langle V_p|V_k\rangle$ in 
(\ref{hratio}) is replaced by $\langle V_p(l)|V_k(r)\rangle$ (in the case of the double 
projection; for the single projection there is no overlap). 
The acceptance rate of a state update is similar to that of an operator string update with
a small $R$, and often we find it advantageous to combine state and operator updates.
To save some time, one can tentatively accept/reject a bond update based solely on an 
amplitude ratio---Eq.~(\ref{hratio}) without the overlaps---and then calculate the overlap 
and the propagation weight for a final accept/reject probability only for tentatively accepted 
bond updates.

In the variational calculation $\langle H\rangle$ is minimized with respect to all
$h(x,y)$. With a recently developed stochastic optimization method \cite{jielou}, all the $\propto N$ 
amplitudes can be minimized for moderate-size lattices (up to $32\times 32$ sites were  
considered in Ref.~\cite{jielou}). In principle we could follow Liang \cite{lia90} and use the best 
possible variational state as our trial state in the projector QMC method---indeed this can be expected 
to be the optimum starting point. However, we will now describe a scheme which delivers a trial state 
nearly as good as the best variational state, at a smaller computational cost.

Consider the probability distribution $P(x,y)$ of valence bonds. In an amplitude-product state
$(\ref{psihprod})$ we would have $P(x,y) \sim h(x,y)$, were it not for the ``hard-core" constraint
of only one bond per site. Even with this constraint, it is clear that the probabilities and amplitudes
are related in a monotonic way; increasing $h(x,y)$ for some given $(x,y)$, while keeping the
other amplitudes fixed, will lead to a larger $P(x,y)$. This fact can be exploited in constructing
a good trial state. We define two different probability distributions, $P_0(x,y)$ and $P_m(x,y)$, 
the former being the just discussed bond probability in a trial state of the form (\ref{psihprod}) 
and the latter the probability distribution in the projected state. For sufficiently large $m$, 
$P_m$ is an exact property of the ground state, whereas $P_0$ is a property of the trial state and 
is in general different from $P_m$. However, for given $m$, we can adjust the amplitudes $h(x,y)$ 
of the trial state such that $P_m(x,y)=P_0(x,y)$ for all $x,y$. If this is done for $m$ sufficiently 
large, then our trial state has a bond distribution identical to that of the exact ground state. 
Such a state is often almost as good as the best variational state. 
The reason that this is useful in practice is that it is very easy to adjust the amplitudes to achieve 
self-consistency. Because of the monotonous relationship between $h(x,y)$ and $P_0(x,y)$, we can 
simply increase $h(x,y)$ by some amount if $P_0(x,y) < P_m(x,y)$ and decrease it if 
$P_0(x,y) > P_m(x,y)$, and repeat this until self-consistency is achieved. We use the following 
scheme to update the amplitudes after the $k$th iteration;
\begin{equation}
\ln[h(x,y)] \to \ln[h(x,y)] + {\rm RAN}\cdot \beta(k) \cdot {\rm sign}[P_m(x,y) - P_0(x,y)],
\label{hupdate}
\end{equation}
where ${\rm RAN}$ is a random number in the range $[0,1)$ and $\beta(k)$ decreases with
the iteration step $k=1,2,\ldots$, according to $\beta(k) \propto k^{-\alpha}$. For the exponent, 
we typically use $\alpha=3/4$. To evaluate the probabilities $P_0$ and $P_m$ 
(in two independent simulations; 
with and without projection of the trial state), the number of Monte Carlo sweeps does not have 
to be very large, because we only need the sign of the difference of the two probabilities.
We normally use on the order of 100-1000 sweeps per iteration. Even if the stochastically evaluated
sign in (\ref{hupdate}) occasionally may be wrong, it is correct on average and the amplitudes 
typically converge to a self-consistent solution after a few hundred iterations. Due to the 
stochastic nature of the procedure, self-consistency of course obtains only to within some 
statistical error, which can be reduced by increasing the number iterations and/or 
sweeps per iteration.

\begin{figure}[t]
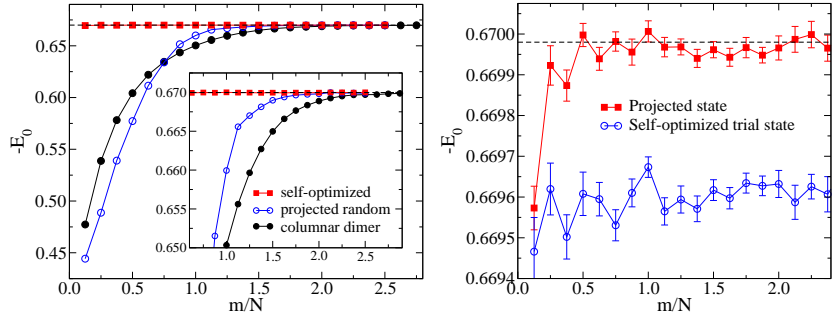

\centerline{
\includegraphics[width=5.5cm]{econv.eps}~
\includegraphics[width=5.22cm]{eoptim.eps}}
\caption{Left: Energy versus projection length for a $16\times16$ system, 
using three different trial states (labeled in the inset). Right: The energy of 
the self-optimized trial state compared with the projected energy using that trial state.
The dashed line shows the energy evaluated independently using the SSE method \cite{energy2d}}
\label{econv}
\end{figure}

Fig.~\ref{econv} shows results for the energy of a $16\times 16$ system obtained in double 
projections with three different trial states; a columnar dimer state, a randomly picked state 
generated while projecting the columnar state, as well as the self-optimized state. Already 
for the shortest projection, $m=N/8=32$, the self-optimized state gives a projected energy which 
deviates by only $0.06$\% from the exact ground state energy, and for larger $m$ the energy is 
exact within statistical errors. The other two trial states also lead to the correct energy but only 
for much larger $m$. The energy of the self-optimized trial state itself is shown in the 
right panel---its error is as small as that of the best variational amplitude-product state \cite{jielou}.

\begin{figure}[t]
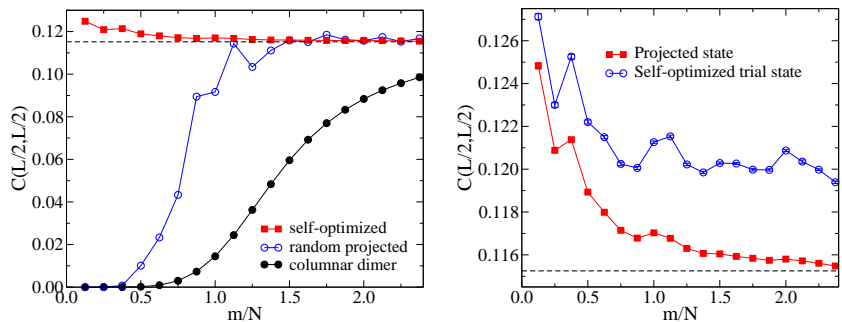

\centerline{
\includegraphics[width=5.5cm]{sconv.eps}~~~
\includegraphics[width=5.1cm]{soptim.eps}}
\caption{Long-distance spin correlations in the same runs as in Fig.~\ref{econv}. A different 
``random projected" state is picked for each $m$, which leads to an un-smooth curve. Fluctuations 
beyond the small statistical errors in the right panel reflect differences in how closely the 
rather short self-optimization runs have approached $P_0=P_m$}
\label{sconv}
\end{figure}

Fig.~\ref{sconv} shows the long-distance spin correlation calculated in the same runs. Again, the
self-optimized state delivers superior results, although here the convergence is not as fast as
for the energy. The error of the spin correlation in the trial state (right panel) is about $4$\% for
large $m$, which is twice the error in the best variational state \cite{jielou}. Thus, the self-optimized
state is not identical to the best variational state, and an even faster convergence 
could be achieved by using the fully optimized variational state. However, the variational 
calculation is much more time consuming than the self-optimization.

We can go beyond the amplitude-product state by taking into account bond correlations. We are 
currently exploring this with both variational and self-optimized states.

\section{Triplet excitations}

A unique advantage of the VB basis is that an $m_z=0$ triplet state can be projected simultaneously 
with the singlet, with essentially no additional overhead. Any triplet can be expanded in VB states
where one of the bonds corresponds to a triplet \cite{hul38}; $(i,j) \to [i,j]$, where
\begin{equation}
[i,j]=(|\uparrow_i\downarrow_j\rangle+|\downarrow_i\uparrow_j\rangle)/\sqrt{2}.
\end{equation}
Formally, such a triplet can be created by acting on a singlet with $S^z_i-S^z_j$;
\begin{equation}
(S^z_i-S^z_j)(i,j)=[i,j].
\end{equation}
To create a triplet $|\tau({\bf q})\rangle$ with some momentum ${\bf q}$, we can 
apply 
\begin{equation}
S^z_{\bf q} = \frac{1}{\sqrt{N}}\sum_j {\rm e}^{i{\bf q}\cdot {\bf r}_j} S^z_j
\label{szq}
\end{equation}
to a singlet state $|\sigma (0)\rangle$ with zero momentum;
\begin{equation}
S^z_{\bf q}|\sigma(0)\rangle \propto |\tau({\bf q})\rangle.
\end{equation}
The amplitude-product state (\ref{psihprod}) for a periodic-boundary system has $q=0$ if the 
number of bonds, $N/2$, is even, whereas for odd $N/2$ it has ${\bf q}=(\pi,\pi)$. This simply follows 
from the fact that sublattice $A \to B$ and $B \to A$ when translating by one lattice constant, whence
each singlet $(\ref{singletbond})$ acquires a minus sign. We typically work with systems with even
$N/2$ (e.g., $L\times L$ lattices with even $L$) and so we will here consider $q=0$ singlets.

At the antiferromagnetic wave-vector, ${\bf Q}=(\pi,\pi)$, $S^z_{\bf Q}$ acting on an arbitrary
VB basis state $|V\rangle$ can be written as a sum of $N/2$ terms of the form $(S^z_i-S^z_j)|V\rangle$, 
with $i,j$ corresponding to the sites connected by bonds. Operating on a $q=0$ singlet thus gives
\begin{eqnarray}
S^z_{\bf Q}|\sigma(0)\rangle & = & S^z_{\bf Q} \sum_k f_k |(i_1,j_1)_k (i_2,j_2)_k \cdots
(i_{N/2},j_{N/2})_k\rangle \nonumber \\
& = & \sum_{b=1}^{N/2} \sum_k f_k |(i_1,j_1)_k \cdots [i_b,j_b]_k \cdots
(i_{N/2},j_{N/2})_k\rangle ,
\end{eqnarray}
where the unspecified coefficients $f_k$, e.g., the amplitude products in (\ref{psihprod}),
have translational invariance built in. Thus, for a triplet with ${\bf q}={\bf Q}$ the wave 
function phases are buried in the definition of the singlets. Often, the lowest excitation
of a Heisenberg system is a ${\bf q}={\bf Q}$ triplet, which we thus can sample without any
difficulties with signs or phases. We consider this case first, before turning to triplets
with arbitrary momentum.

There are two possible actions of a singlet projector $H_{ij}$ on a triplet bond;
\begin{eqnarray}
& & H_{ij}|\cdots [i,j]\cdots \rangle = 0, \label{tdia} \\
& & H_{ij}|\cdots [i,k]\cdots (l,j)\cdots \rangle =
\hbox{$\frac{1}{2}$}|\cdots (i,j)\cdots [l,k]\cdots \rangle, \label{toff}
\end{eqnarray}
i.e., a diagonal operation on a triplet bond destroys the state whereas an off-diagonal operation on 
one triplet and one singlet bond creates a singlet at the sites on which the operator acts and moves 
the triplet to the other two sites involved. Importantly, the matrix element remains the same as in the 
off-diagonal operation on two singlet bonds. The weight of a triplet path is therefore the same 
as the corresponding singlet path (\ref{wrk}), except that the triplet dies (giving zero weight)
if an operator in $P_r$ acts diagonally on the triplet bond. We can thus measure triplet properties
using paths generated in a singlet simulation, by considering only those triplet paths that do survive 
the propagation. In the trial state we have $N/2$ possible locations of the triplet. All of them can 
be attempted collectively in a single propagation, by keeping counters $t(i)$ for the number of 
surviving states in which the triplet is connected to site $i$ (with $i$ on sublattice A). 
Initially $t(i)=1$ for all $i$. During the propagation, for each diagonal operation (\ref{tdia}) 
$t(i) \to t(i)-1$, and for each off-diagonal operation (\ref{toff}) $t(i) \to t(i)-1$, 
$t(l) \to t(l)+1$. Eventually, as $m \to \infty$, all triplets die; $t(i)=0$ for all $i$, but 
typically there are enough survivors left at large enough $m$ to compute converged triplet properties. 

An added advantage of calculating singlet and triplet properties in the same run is that there
are error cancellations which in some cases can increase the statistical precision of differences, 
e.g., the singlet-triplet gap \cite{sandvikvb}, 
\begin{equation}
\Delta=E_T(\pi,\pi)-E_S(0,0),
\label{gap}
\end{equation}
by up to orders of magnitude relative to two independent calculations. The triplet
energy $E_T(\pi,\pi)$ can be estimated using (\ref{esingle}), taking into account that a
diagonal operation on a triplet bond gives zero, i.e., $n_d \to n_d - n_t$, where, for
surviving triplet configurations, $n_t=0,1$ is the number of triplet bonds of length $1$.
Other triplet properties have been discussed in Ref.~\cite{kevinvb}. 

We now discuss calculations with triplets of arbitrary momentum ${\bf q}$. The energy can
be evaluated according to
\begin{equation}
E({\bf q}) = \frac
{\sum_r \langle \sigma (0)|S^z_{-{\bf q}} H P_r S^z_{\bf q}|\sigma(0)\rangle}
{\sum_r \langle \sigma (0)|S^z_{-{\bf q}} P_r S^z_{\bf q}|\sigma(0)\rangle}.
\label{eq1}
\end{equation}
We want to evaluate this expression using the sampled $q=0$ singlet bond configurations, 
and so we rewrite it as
\begin{equation}
E({\bf q}) = \frac
{\sum_r \langle \sigma (0)|S^z_{-{\bf q}} H P_r S^z_{\bf q}|\sigma(0)\rangle}
{\sum_r \langle \sigma (0)| P_r |\sigma(0)\rangle}
\left ( \frac
{\sum_r \langle \sigma (0)|S^z_{-{\bf q}} P_r S^z_{\bf q}|\sigma(0)\rangle}
{\sum_r \langle \sigma (0)| P_r |\sigma(0)\rangle} \right ) ^{-1}.
\label{eq2}
\end{equation}
The two factors can be evaluated based on sampling the propagations $P_r$ and the 
amplitude-product state $|\sigma(0)\rangle$ [which we have not explicitly written as
a sum of bond configurations in (\ref{eq1}) and (\ref{eq2})]. In the singlet energy 
(\ref{hexp}), we could pick the N\'eel state for $\langle \Psi |$ and then obtained 
the very simple expression (\ref{esingle}). Now we must consider the overlap 
with a momentum ${\bf q}$ triplet state. We use $\langle \sigma (0)|S^z_{-{\bf q}}$, 
but in (\ref{eq2}) have rewritten $E({\bf q})$ so that only the overlap with $\langle \sigma (0)|$
has to be considered for the sampling weight. Phases arising from $S^z_{\bf q}$ only appear in 
the measurements, but in the end we have to evaluate the ratio of the two quantities in 
(\ref{eq2}), which can be challenging in practice. However, close to ${\bf q}=(\pi,\pi)$ and 
$(0,0)$ [${\bf q} \not= (0,0)$] we find  that it can be done; in some cases the method works 
even far away from these momenta.

\begin{figure}[t]
\centerline{\includegraphics[width=10cm]{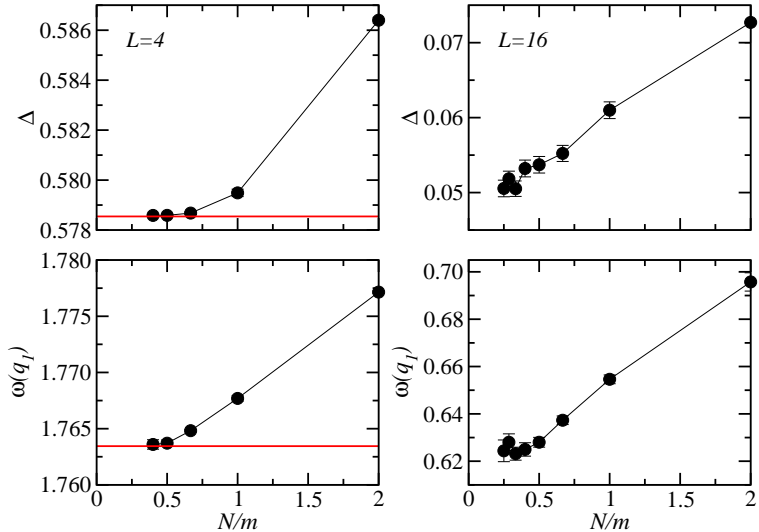}}
\caption{Projection-length convergence of the singlet-triplet gap and the excitation energy at 
momentum ${\bf q}_1 = (\pi-2\pi/L,\pi)$ for $L=4$ (left) and $16$ (right) lattices. The horizontal 
lines show the exact results for $L=4$}
\label{eq}
\end{figure}

We define the dispersion relative to the gap (\ref{gap}) at $(\pi,\pi)$;
\begin{equation}
\omega ({\bf q}) = E_T({\bf q}) - \Delta.
\end{equation}
Fig.~\ref{eq} shows results for $\Delta$ and $\omega({\bf q}_1)$, where ${\bf q}_1$ is the momentum
closest to but not equal to $(\pi,\pi)$; ${\bf q}_1 = (\pi-2\pi/L,\pi)$. We show the convergence
as a function of $N/m$ for  $4\times 4$ and  $16\times 16$ lattices, comparing with exact
diagonalization results in the former case. From (\ref{projection}) one  would expect the
convergence to be asymptotically exponential, which is seen clearly for $L=4$. For $L=16$, the 
three largest-$m$ points are equal within statistical errors, suggesting
that these results are also close to converged. Using $\omega ({\bf q}_1)=0.62$ for the
$L=16$ system gives the spin-wave velocity $c=1.58$, in very close agreement with the known 
value \cite{energy2d}.

Another useful quantity accessible with the VB projector is the matrix element 
$\langle \tau ({\bf q})|S^z_{\bf q}|\sigma(0)\rangle$, the square of which gives the single-magnon
weight in the dynamic structure factor (which is experimentally measurable using neutron 
scattering). It can be calculated in a way similar to $E({\bf q})$, and we have done so 
successfully for ${\bf q}$ close to  $(\pi,\pi)$. Results will be presented elsewhere.

Finally, we also note that the triplet bond-length distribution gives a direct, albeit
basis dependent, window into the ``spinon" aspects \cite{sachdevrmp} of the excitations. Spinon 
deconfinement should be manifested as a delocalized distribution function, whereas two spinons bound 
into a magnon or ``triplon" should be reflected in a well-defined peak in the distribution function.
We are currently exploring this.

\section*{Acknowledgments}

We would like to thank H. G. Evertz for useful discussions. This work was supported by the National 
Science Foundation under grant No.~DMR-0513930.

%INDEX%%%%%%%%%%%%%%%%%%%%%%%%%%%%%%%%%%%%%%%%%%%%%%%%%%%%%%%%%%%%%%%
% Please check with the editor of your book whether he plans to
% include a "mutual" subject index - if so, please code your entries
% in the standard syntax. For your own purposes you may print your
% "personal" index by using the following commands:
%
%\clearpage
%\addcontentsline{toc}{section}{Index}
%\flushbottom
%\printindex
%%%%%%%%%%%%%%%%%%%%%%%%%%%%%%%%%%%%%%%%%%%%%%%%%%%%%%%%%%%%%%%%%%%%%

\end{document}